\documentclass{article}
\usepackage{spconf,amsmath,graphicx,bm}
\usepackage{subcaption}
\usepackage{amssymb}
\usepackage{mathtools} 
\usepackage[bottom]{footmisc}
\usepackage{breqn}
\usepackage{threeparttable}
\usepackage[square,numbers,sort&compress]{natbib}
\usepackage{url}
\usepackage{algorithm}
\usepackage{multirow} 
\usepackage[normalem]{ulem}  
\usepackage{etoolbox}
\newcommand{\ubold}{\fontseries{b}\selectfont}
\robustify\ubold

\usepackage{siunitx}
\def\x{{\mathbf x}}
\def\y{{\mathbf y}}
\def\q{{\mathbf q}}
\def\r{{\mathbf r}}
\def\A{{\mathbf A}}

\def\H{{\mathbf H}}
\def\R{{\mathbf R}}
\def\Q{{\mathbf Q}}

\def\G{{\mathbf G}}

\def\z{{\mathbf z}}

\def\S{{\mathbf S}}
\def\Id{{\textbf{Id}}}

\newcommand{\Sigmab}{{\bm \Sigma}}
\newcommand{\Omegab}{{\bm \Omega}}
\newcommand{\Deltab}{{\bm \Delta}}
\newcommand{\Phib}{{\bm \Phi}}
\newcommand{\Psib}{{\bm \Psi}}
\newcommand{\mub}{\boldsymbol\mu}

\usepackage{color}

\def\ctA{{\mathrm{ct}_{/\mathbf{A}}}}

\title{GraphIT: Iterative reweighted $\ell_1$ algorithm for sparse graph inference in state-space models}
\name{\'Emilie Chouzenoux$^{1}$ and V\'ictor Elvira$^{2}$}

\address{$^{1}$ Universit\'e Paris-Saclay, CentraleSup\'elec, Inria, CVN, Gif-sur-Yvette, France \\
    $^{2}$ School of Mathematics, University of Edinburgh, United Kingdom
    \thanks{V.E. acknowledges support from the Agence Nationale de la
Recherche of France under PISCES (ANR-17-CE40-0031-01) project, the Leverhulme Research Fellowship (RF-2021-593), and by ARL/ARO under grant W911NF-22-1-0235. E.C. acknowledges support from the European Research Council under Starting Grant MAJORIS ERC-2019-STG-850925. }}
    
\begin{document} 
\maketitle
\begin{abstract} 
State-space models (SSMs) are a common tool for modeling multi-variate discrete-time signals. The linear-Gaussian (LG) SSM is widely applied as it allows for a closed-form solution at inference, if the model parameters are known. However, they are rarely available in real-world problems and must be estimated. Promoting sparsity of these parameters favours both interpretability and tractable inference. In this work, we propose GraphIT, a majorization-minimization (MM) algorithm for estimating the linear operator in the state equation of an LG-SSM under sparse prior. A versatile family of non-convex regularization potentials is proposed. The MM method relies on tools inherited from the expectation-maximization methodology and the iterated reweighted-l1 approach. In particular, we derive a suitable convex upper bound for the objective function, that we then minimize using a proximal splitting algorithm. Numerical experiments illustrate the benefits of the proposed inference technique. 
\end{abstract}
\begin{keywords}
State-space model, graphical inference, sparsity, iterative reweighted scheme, MM algorithm.
\end{keywords}
\section{Introduction}
\label{sec:intro} 

Signal processing applications often involve dealing with discrete-time multi-variate series that need to be processed to perform statistical tasks such as prediction. 
State-space models (SSMs) relate the observation process with a hidden state that evolves over time. 
This versatile approach allows to describe complex phenomena through the dynamics of the hidden state.
It also allows for the incorporation physical knowledge about the system and to retrieve relevant structure that can be inferred when processing the data. 
There exists abundant literature in graphical models for multi-variate time series, e.g., \cite{Eichler2012,Bach04,Barber10,ioannidis2019semi}, and more recently also within SSMs \cite{Chouzenoux2020,ElviraGraphEM2022}, including also fully probabilistic approaches \cite{cox2022parameter,cox2023sparse}. 
There are multiple applications of such graphical representations, e.g., in  biology \cite{Pirayre,Luengo19}, social network analysis \cite{Ravazzi}, or neuroscience \cite{Richiardi}. 

The graphical approach in SSMs benefits from the promotion of sparsity with, at least, the following three advantages. First, sparsity enhances interpretability, which is highly desired for instance in models where the hidden state has a physical interpretation. Second, it reduces the effective dimension of the parameters, alleviating the curse of dimensionality during estimation. 
%
%
Third, it is possible to promote other properties that are complementary with sparsity, allowing to introduce useful prior knowledge on the system, e.g., stability of the hidden process and spectral constraints \cite{chiu2021lowrank}.

{In \cite{Chouzenoux2020,ElviraGraphEM2022}, we introduced GraphEM, an expectation-maximization (EM) method for parameter estimation in linear Gaussian (LG) SSMs. Specifically, we addressed the challenging task of estimating the transition matrix that encodes the Markovian dependencies in the evolution of the multi-variate state. GraphEM is based on a novel perspective that relates the transition matrix to the adjacency matrix of a directed graph, encoding causal relationship among state dimensions in the Granger-causality sense. The M-step in the EM procedure relies on a proximal splitting, benefiting from sound convergence guarantees. {GraphEM thus extends the well-established EM technique for maximum likelihood estimation of LG-SSM parameters~\cite{Zoweis,Ghahramani,shumway1982approach,Cappe2005} into a maximum a posteriori estimator that suitably encompasses sophisticated prior terms.} However, it is limited to convex regularizers.
{In the literature of sparse graph inference, non-convex penalties such as SCAD \cite{Fan2001}, MCP \cite{Zhang2010}, CEL0 \cite{Soubies2015} have shown to be more suited than their convex surrogate $\ell_1$ norm (i.e., Lasso) to properly reconstruct very sparse graphs~\cite{Ying2020,Cardoso2021,WilliamsSurvey}. These penalties can be viewed as continuous non-convex approximations to the quasi-norm $\ell_0$. However, non-convexity prevents the use of standard optimization tools. An efficient strategy is to resort to iterative reweighted (IR) schemes~\cite{Wipf2010,Huber}, 
which are based on the majorization-minimization (MM) principle \cite{Sun2017} to recast the complicated objective function into a sequence of tractable upper bounds. 
IR$\ell_1$ algorithms have been widely applied to several signal processing problems \cite{Daubechies2009,Chartrand,Ochs2015,Carrillo2012,Sun2018,Ying2020b}, but, to our knowledge, remain unstudied within SSMs.}
%

{In this paper, we propose a novel method called GraphIT that combines the EM technique with the powerful IR$\ell_1$ strategy for the estimation of model parameters in SSMs.  More precisely, GraphIT is able to provide an estimation of the LG-SSM transition matrix for a wide class of non-convex sparsity enhancing priors. Thanks to the IR$\ell_1$ framework, GraphIT algorithm benefits from simple and not expensive updates. Experimental results on several synthetic datasets allow us to assess the ability of our method for improving the recovery rate for sparse transition matrices.}

The paper is structured as follows. In Section \ref{sec:format}, we describe the model, the filtering/smoothing algorithms, and revisit the GraphEM framework. The novel GraphIT algorithm is presented in Section \ref{sec:proposed}. The paper ends with numerical results in Section \ref{sec:experiment} and some concluding remarks in Section~\ref{sec:conclusion}.

\section{BACKGROUND}
\label{sec:format}

\subsection{Inference in linear state-space model}
We consider the LG-SSM described by
 \begin{equation}
   \begin{array}{ll}
       \x_{k} &= \A \x_{k-1} + \q_{k},   \\
     \y_k &= \H \x_k + \r_{k},
    \end{array}
    \label{eq:model}
    \end{equation}
for $k=1,\ldots,K$, where $\x_{k}\in \mathbb{R}^{N_x}$ and $\y_{k}\in \mathbb{R}^{N_y}$ are the hidden state and observation at time $k$, respectively, $\A = \mathbb{R}^{N_x \times N_x}$, $\H = \mathbb{R}^{N_y \times N_x}$, and $\{ \q_k \}_{k=1}^K \sim \mathcal{N}(0,\Q)$ and $\{ \r_k \}_{k=1}^K \sim \mathcal{N}(0,\R)$ the i.i.d. state and observation noise processes, respectively. The distribution of the initial state is $\x_0 \sim \mathcal{N}(\x_0 ; \mub_0, \Sigmab_0)$ with known $\mub_0$ and $\Sigmab_0$.

In SSMs, the filtering problem consists in the computation of the so-called filtering distributions, $p(\x_k|\y_{1:k})$, where  $\y_{1:k} \triangleq \{ \y_j \}_{j=1}^k$. 
In the LG-SSM of  Eq. \eqref{eq:model}, the  Kalman filter (KF) exactly computes the filtering distributions in a sequential and efficient manner, yielding $p(\x_k|\y_{1:k}) = \mathcal{N}(\x_k|\mub_k,\Sigmab_k)$, i.e., for every $k=1,\ldots,K$, it computes the mean $\mub_k$ and covariance $\Sigmab_k$ of each Gaussian filtering distribution ~\cite{Kalman60}. 
The smoothing problem aims at computing $p(\x_k|\y_{1:K})$, i.e., the posterior probability density functions (pdfs) conditioning on the whole set of observations instead, for all $k=1,\ldots,K$. In the  LG-SSM, the smoothing pdfs are also available in a closed-form expression (again, Gaussian pdfs) by the {Rauch-Tung-Striebel (RTS)} smoother, $p(\x_k|\y_{1:K}) = \mathcal{N}(\x_k|\mub_k^s,\Sigmab_k^s)$. Both KF and RTS smoother require to know the model parameters {$\mub_0$, $\Sigmab_0$}, $\A$, $\H$, $\Q$, and~$\R$. The estimation of the model parameters in the model above has been a subject of intense research, e.g., recently \cite{sharma2021recurrent,sharma2020blind} (see \cite[Ch. 12]{Sarkka} for a review).

 \subsection{GraphEM to estimate the transition matrix}

GraphEM \cite{ElviraGraphEM2022,Chouzenoux2020} is an EM-based algorithm that we recently introduced for the maximum a posteriori (MAP) estimation of $\A$ under log-concave prior distribution $p(\A)$. GraphEM aims at finding the minimum of $\mathcal{L}(\A) \triangleq \mathcal{L}_0(\A) + \mathcal{L}_{1:K}(\A)$ with $\mathcal{L}_0(\A) \triangleq - \log p(\A)$ and $\mathcal{L}_{1:K}(\A) \triangleq - \log p(\y_{1:K}| \A)$. The matrix $\A$ is interpreted in GraphEM to encode a directed graph among the state dimensions. Prior $ p(\A)$ can encode structural information on such graph (e.g., sparsity). This graph plays an important role in the SSM interpretation, with links to Granger-causality models (see a discussion in \cite{ElviraGraphEM2022}). The likelihood of $\A$ associated to the model in Eq. \eqref{eq:model} is
\begin{align}
   &  \mathcal{L}_{1:K}(\A) = \sum_{k=1}^K \tfrac{1}{2} \log | 2 \pi \S_k| + \frac{1}{2} \z_k^\top \S_k^{-1} \z_k,
\end{align}
where $\z_k = \y_k - \H\A\mub_{k-1}$  and $\S_k$ is the covariance matrix of the predictive distribution $p(\y_{k}|\y_{1:k-1}, \A)= \mathcal{N}\left(\y_{k};\H\A\mub_{k-1},\S_{k}\right)$, both being obtained by the KF, run for a given $\A$ (see \cite[Section 4.3]{Sarkka}). The direct minimization of $\mathcal{L}$ is difficult due to (i) the implicit form of $\mathcal{L}_{1:K}$, and (ii) the possibly non-differentiability of $\mathcal{L}_0$. The GraphEM approach implements an EM strategy, generalizing the one in \cite{shumway1982approach}, by relying on successive upper bounds of $\mathcal{L}_{1:K}$, that leads to obtain a sequence of {tractable} inner problems. 

Let us consider KF/RTS outputs for a given $\A' \in \mathbb{R}^{N_x \times N_x}$. Denote, for every $k \in\{1,\ldots,K\}$, $\G_k = \Sigmab_k (\A')^\top (\A' \Sigmab_k (\A')^\top + \Q)^{-1}$, and set $\Psib = \sum_{k=1}^K \Sigmab_k^s + \mub_k^s (\mub_k^s)^\top$, $\Phib  =  \sum_{k=1}^K \Sigmab_{k-1}^s + \mub_{k-1}^s (\mub_{k-1}^s)^\top$, and $\Deltab =   \sum_{k=1}^K \Sigmab_k^s \G_{k-1}^\top + \mub_k^s (\mub_{k-1}^s)^\top$. Then, according to \cite[Chap. 12]{Sarkka}, 
\begin{align}
& (\forall \A \in \mathbb{R}^{N_x \times N_x}) \quad \mathcal{L}_{1:K}(\A)  \leq  \nonumber \\
& - \int p(\x_{0:K} | \y_{1:K},\A') \log p(\x_{0:K},\y_{1:K} | \A) \rm{d} \x_{0:K} + \ctA \nonumber \\
& \; = \tfrac{1}{2}  \text{tr} \left(\Q^{-1} (\Sigmab - \Deltab \A^\top - \A \Deltab^\top + \A \Phib \A^\top) \right) + \ctA,
   \label{eq:majQ}
\end{align}
where $\rm{tr}$ is the trace operator, and $\ctA$ is a constant term not depending on~$\A$, such that equality holds at $\A = \A'$.   
The GraphEM algorithm iterates alternating between an E-step, building the right term in~\eqref{eq:majQ}, and an M-step minimizing the sum of this term with $\mathcal{L}_0$, using a proximal splitting scheme. In the simple case of a Laplace prior (i.e., $\ell_1$ norm penalty), the minimization in the M-step can be performed by a Douglas-Rachford algorithm, leading to the simpler form of GraphEM in \cite{Chouzenoux2020}, while a more sophisticated primal-dual splitting technique is proposed in \cite{ElviraGraphEM2022} for dealing with a generic convex penalty. Convergence guarantees for the resulting GraphEM iteration are discussed in \cite{ElviraGraphEM2022}.

\vspace{-2mm}

\section{PROPOSED GRAPHIT APPROACH}
\label{sec:proposed}

\subsection{Considered class of penalties}
Our study focuses on the class of non-convex sparsifying penalties that have been studied for instance in \cite{Gasso2009} (see also \cite{PanIRL1} in the context of robust estimation). We define 
\begin{equation}
\mathcal{L}_0(\A) =  \sum_{1 \leq i,j \leq N_x} \rho({|A_{i,j}|}), \label{prior}
\end{equation}
where $A_{i,j}$ denotes the $(i,j)$ entry of $\A$, and $\rho: \mathbb{R} \to [0,+\infty)$ is a potential satisfying:  
\begin{enumerate}
    \item $\rho$ is continuous, and it is increasing on $[0,+\infty)$ with $\rho(0) = 0$;
    \item $\rho$ is differentiable on $(0,+\infty)$, with derivative (when defined) denoted $\rho'$;
    \item $\rho'$ is decreasing on $(0,+\infty)$ and $\lim_{u \to 0^+} \rho'(u) = \gamma \in (0,+\infty)$.
\end{enumerate}
A large class of potential functions satisfy the above requirements. Some relevant examples are listed in Table \ref{tableprior}. Except for SCAD, in our examples, epi-convergence of $u \mapsto \rho(|u|)$ to the discrete potential $u \mapsto \delta_{u \neq 0}$ (equals $0$ for a zero-valued input; $1$ otherwise) is established for $\lambda \to 0^+$. Then, the prior in \eqref{prior} can be seen as an approximation of $\gamma \ell_0$.  Fig. \ref{fig:penalty} shows the plots for a subset of potentials, with $\gamma = 1$. We also show $\ell_1$ and $\ell_0$ penalties for comparison. The considered~class of priors is continuous, but still non-convex approximations of~$\ell_0$.

\begin{table}
\centering
\scalebox{0.6}{
\renewcommand{\arraystretch}{2}
\begin{tabular}{|c|c|c|}
\hline
Name & $\rho(|u|)$ & $\rho'(|u|)$\\
\hline
\hline
log-sum & $\gamma \lambda (\log(|u| + \lambda) - \log(\lambda)), \,\text{with } \lambda>0$ & $\gamma \lambda (|u| + \lambda)^{-1}$\\
\hline
atan & $\frac{\gamma}{\lambda}\, \text{atan}(\lambda |u|), \, \text{with }\lambda>0$ &  $\frac{\gamma}{1 + \lambda^2 u^2}$ \\
\hline
Mangasarian & $\frac{\gamma}{\lambda}(1 - \exp(-\lambda |u|))$ & $\gamma \exp(-\lambda |u|)$ \\
\hline
MCP & $\begin{cases} \gamma |u| - \frac{u^2}{2 \lambda} & \text{if }|u|\leq \lambda \gamma\\
\frac{\lambda \gamma^2}{2} & \text{otherwise} \end{cases},\text{with }\lambda>0$ & $\begin{cases} \gamma - \frac{|u|}{\lambda} & \text{if }|u|\leq \lambda \gamma\\
0 & \text{otherwise} \end{cases}$\\
\hline
SCAD & $\begin{cases} \gamma |u| & \text{if }|u|\leq \gamma\\
- \frac{\gamma^2 - 2 a \gamma |u| + u^2}{2(a-1)} & \text{if }\gamma<|u|\leq a \gamma\\
\frac{(a+1)\gamma^2}{2} & \text{otherwise} \end{cases},\text{with }a>2$ & $\begin{cases} \gamma & \text{if }|u|\leq \gamma\\
\gamma - \frac{2 |u| - (2 a -1)\gamma }{2(a-1)} & \text{if }\gamma<|u|\leq a \gamma\\
0 & \text{otherwise} \end{cases}$ \\
\hline
\end{tabular}    
}                                                          
\caption{Example of potential functions and their derivatives with hyper-parameter $\gamma>0$ such that $\rho'(0^+) = \gamma$.} \label{tableprior}
\end{table}

\subsection{IR$\ell_1$ approach}
 
The properties assumed on the potential $\rho$ yield a key property, that is at the core of the so-called IR$\ell_1$ optimization approach \cite{Zou2008}. Let $v \in \mathbb{R}$. Then, 
\begin{equation}
(\forall u \in \mathbb{R}) \quad \rho(|u|) \leq \rho'(|v|)(|u| - |v|) + \rho(|v|),
\end{equation}
where equality holds at $u = v$. 
A proof for this result can be found in \cite{Ochs2015}. The following convex upper bound is then obtained, for~\eqref{prior}, at some $\A' \in \mathbb{R}^{N_x \times N_x}$:
\begin{equation}
\mathcal{L}_0(\A) \leq \| \Omegab(\A') \odot \A \|_1 + \ctA, \label{priorw}
\end{equation}
where $\ctA$ is such that equality holds at $\A = \A'$. Hereabove, $\Omegab(\A') \in [0,+\infty)^{N_x \times N_x}$ is a matrix of weights defined as $\Omega_{i,j} = \rho'(|A_{i,j}|)$ for every $(i,j)$. Moreover, $\odot$ is the Hadamard (i.e., element-wise) product. We provide in Table \ref{tableprior} the expression for {$\rho'(|\cdot|)$} for our examples. The IR$\ell_1$ technique amounts at minimizing $\mathcal{L}_0 + f$, where $f$ a simple function (typically a quadratic term), by iteratively solving the tractable problem of the minimization of \hbox{$\| \Omegab(\A') \odot \cdot \|_1 + f$}. 
When $f$ is a least square discrepancy, the inner problem in IR$\ell_1$ can be solved using one of the several efficient approaches proposed in the literature for penalized least squares under Lasso penalty \cite{Combettes2021,Daubechies,Beck}. 

\begin{figure}
\centering
\includegraphics[width = 0.42\textwidth]{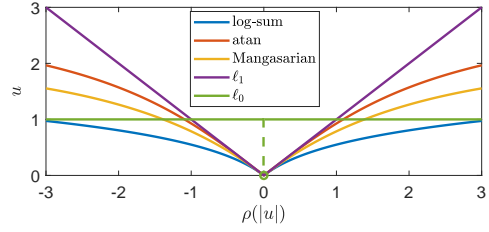}
\caption{Example of functions $\rho(|\cdot|)$, with $\lambda = 0.5$ and $\gamma = 1$.}
\label{fig:penalty}
\end{figure}

\vspace{-2mm}
\subsection{GraphIT algorithm}

In the following, we describe the GraphIT algorithm, which is summarized in Algorithm~\ref{algo:GraphIT}. GraphIT is an MM algorithm, to search for the MAP estimate of $\A$ within our SSM (i.e., the minimizer of $\mathcal{L}$), under the prior defined in \eqref{prior}. It has an iterative structure, which alternates over majorization and minimization steps. The difficulty lies in the construction of the majorization approximation and in the required optimization algorithm to minimize it.

\begin{algorithm}[h!]
{\footnotesize
\noindent Initial $\A^{(0)}$ and penalty parameters. Precision $\varepsilon>0$.\\
\noindent For $i = 1,2,\ldots$\\
    $\phantom{a}$\underline{\textit{Majorization:}}\\
        $\phantom{a}$ Run KF and RTS smoother using $\A^{(i-1)}$ and construct
\begin{multline}
(\forall \A \in \mathbb{R}^{N_x \times N_x}) \quad \mathcal{Q}(\A;\A^{(i-1)}) = \\
\!\tfrac{1}{2}\!\text{tr}\!\left(\Q^{-1}\!(\Sigmab\!-\!\Deltab \A^\top\!-\!\A \Deltab^\top\!+\!\A \Phib \A^\top) \right) \!
+ \!\| \Omegab(\A^{(i-1)})\!\odot\!\A \|_1.\!    \label{eq:majFull}
\end{multline} 
        $\phantom{a}$\underline{\textit{Minimization:}}\\
            $\phantom{a}$ Solve $\A^{(i)} = \text{argmin}_{\A} \left( \mathcal{Q}(\A;\A^{(i-1)})\right)$ using DR solver.\\
$\phantom{a}$\underline{\textit{Stopping condition:}}\\         
$\phantom{a}$ If $\|\A^{(i)}-\A^{(i-1)}\|_F \leq \varepsilon \|\A^{(i-1)}\|_F$,  {stop the recursion}.
}
        \caption{GraphIT algorithm}
    \label{algo:GraphIT}
\end{algorithm}

Given an estimation, $\A^{(i-1)}$, coming from the previous iteration, we construct a specific majorization function $\mathcal{Q}(\A;\A^{(i-1)})$ defined in Eq. \eqref{eq:majFull} that combines (i) an EM-based majorization involving outputs from running the KF/RTS with transition matrix $\A^{(i-1)}$, and (ii) a convex upper bound for our prior term in Eq. \eqref{prior} (coming from IR$\ell_1$ approach). 
%
Using \eqref{eq:majQ} and \eqref{priorw}, we can show that the function in Eq. \eqref{eq:majFull} is such that, for every $\A \in \mathbb{R}^{N_x \times N_x}$ and every iteration $i \in \mathbb{N}$, $\mathcal{L}(\A)  \leq  \mathcal{Q}(\A;\A^{(i-1)}) + \ctA$ (heragain, the constant term is set so as to have equality at $\A = \A^{(i-1)}$).
In practice, the minimization of the above upper bound is done through the Douglas-Rachford (DR) algorithm \cite{CombettesDR} initialized at the current iterate $\A^{(i-1)}$. As a virtue of the MM construction, the sequence $\{\mathcal{L}(\A^{(i)})\}_{i \in \mathbb{N}}$ 
is a decreasing sequence converging to a finite limit. 


\section{EXPERIMENTAL EVALUATION}
\label{sec:experiment}


We illustrate the performance of the proposed GraphIT method on a time series  simulated according to model \eqref{eq:model}, with $N_x = N_y \in \{8,16\}$. Sparse matrices $\A$ are randomly generated, with support size (i.e., number of edges in the associated graph) $S \in \{4,8,16\}$. For each of the six scenarios, we randomly select $S$ indexes for which we set the associated entry in $\A$ equals to a realization of an i.i.d. standard normal distribution and with the $N_x^2 - S$ other entries being set to zero. The non-zero entries of $\A$ are scaled so that the spectral norm of $\A$ is less than one to guarantee the stability of the Markov process. We finally set {$\H = \rm{\Id}$}, $K = 10^3$ and $\Q = \sigma_\Q^2 \rm{\Id}$, $\R = \sigma_\R^2 \rm{\Id}$, $\Sigmab_0 = \sigma_0^2 \rm{\Id}$ with $(\sigma_\Q,\sigma_\R,\sigma_0) = (10^{-1},10^{-1},10^{-4})$. 

For each of the six datasets, we ran GraphIT using log-sum penalty {(Table ~\ref{tableprior}, first row)} and precision $\xi = 10^{-3}$ (with a maximum of $50$ iterations). We also display the results of GraphEM using $\ell_1$ penalty, and of the maximum likelihood estimator computed by an EM algorithm (MLEM) {\cite{shumway1982approach,Ghahramani}}. All methods are initialized with a matrix with entries $A_{i,j} = (10^{-1})^{|i-j|}$, rescaled to have spectral norm equals to $0.99$. The methods are compared in terms of relative mean square error (RMSE) on $\A$, accuracy and F1 scores for the graph edge detection, using a threshold of $10^{-10}$ on the entries of matrix $\A$. All results are averaged on $50$ realizations. The regularization parameters of GraphIT and GraphEM are fine-tuned through a grid-search strategy to minimize the RMSE on a single realization.

The results are summarized in Table~\ref{tab:results}. MLEM does not include sparsity prior on $\A$, which explains its poor performance, especially in terms of edge detectability. GraphEM and GraphIT both reach better scores than MLEM. It is noticeable that, the sparser the graph (i.e., the smaller $S$), the better results for GraphIT. The superiority of GraphIT over its competitors on all datasets is remarkable both in terms of RMSE and detection metrics. In particular, despite integrating a sparsifying $\ell_1$ penalty in its estimator, GraphEM struggles for very sparse graphs, leading to an increase in RMSE and a lowered accuracy. In contrast, GraphIT performs well for all metrics. When the level of sparsity decreases (i.e., $S$ increases), both GraphEM and GraphIT tend to perform similarly. 
We also include in the last column of Table~\ref{tab:results}, the averaged computing times over $50$ realizations for each methods. All methods take a similar running time (code is run in Matlab 2021a on an 11th Gen Intel(R) Core(TM) i7-1185G7 3.00GHz with 32 Go RAM).  

{We finally display two examples of graph reconstruction in Fig.~\ref{fig:datasetC}. We compare the true graph (left), with GraphEM (middle) and GraphIT (right) estimates in both setups, i.e., $(N_x,S) = (8,16)$ (top) and $(N_x,S) = (16,16)$ (bottom). We can see the superior ability of GraphIT to recover the graphs shape and weights in the case of two sparse graphs.}

\begin{table}[t]
\caption{Results and computing times for GraphIT, GraphEM, and MLEM.}
\centering
\scalebox{0.8}
{
\begin{tabular}{|c|c||c|c|c|||c|}
\hline
$(N_x,S)$ & method & RMSE & accur.  & F1 & Time (s.)\\
\hline
\multirow{3}{*}{(8,4)} & GraphIT & $\ubold \num{0.18501}$ & $\ubold \num{0.96406}$  & $\ubold \num{0.77566}$ & $\num{1.5232}$\\
  & GraphEM                     & $\num{0.24361}$ & $\num{0.92}$  & $\num{0.6347}$ & $\num{1.1864}$\\
  & MLEM & $\num{0.4008}$ & $\num{0.0625}$  & $\num{0.11765}$ & $\num{1.8619}$\\
\hline
\hline
\multirow{3}{*}{(8,8)} & GraphIT & $\ubold \num{0.16064}$ & $\ubold \num{0.97625}$  & $\ubold \num{0.90108}$ & $\num{1.9074}$\\
  & GraphEM & $\num{0.21361}$ & $\num{0.93625}$ &  $\num{0.78357}$ & $\num{3.4674}$\\
  & MLEM & $\num{0.34493}$ & $\num{0.125}$ &  $\num{0.22222}$ & $\num{1.7744}$\\
\hline
\hline
\multirow{3}{*}{(8,16)} & GraphIT & $\ubold \num{0.19009}$ & $\ubold \num{0.93094}$ & $\ubold \num{0.86055}$ & $\num{1.6625}$\\
  & GraphEM & $\num{0.19428}$ & $\num{0.865}$  & $\num{0.77168}$ & $\num{3.5057}$\\
  & MLEM & $\num{0.24829}$ & $\num{0.25}$ &  $\num{0.4}$ & $\num{2.0396}$\\
\hline
\hline
\multirow{3}{*}{(16,4)} & GraphIT & $\ubold \num{0.23412}$ & $\ubold \num{0.99047}$  & $\ubold \num{0.74901}$ & $\num{4.2184}$\\
  & GraphEM & $\num{0.32516}$ & $\num{0.92797}$  & $\num{0.30182}$ & $\num{4.1853}$\\
  & MLEM & $\num{0.79169}$ & $\num{0.015625}$  & $\num{0.030769}$ & $\num{4.7414}$\\
\hline
\hline
\multirow{3}{*}{(16,8)} & GraphIT & $\ubold \num{0.25705}$ & $\ubold \num{0.98695}$  & $\ubold\num{0.80769}$ & $\num{4.0052}$\\
  & GraphEM & $\num{0.32197}$ & $\num{0.90742}$  & $\num{0.40339}$ & $\num{3.9155}$\\
  & MLEM & $\num{0.64957}$ & $\num{0.03125}$  & $\num{0.060606}$ & $\num{3.9639}$\\
\hline
\hline
\multirow{3}{*}{(16,16)} & GraphIT & $ \ubold {\num{0.34968}}$ & $\ubold \num{0.95883}$ &  $\ubold \num{0.60594}$ & $\num{4.3925}$\\
  & GraphEM & $\num{0.3623}$ & $\num{0.90539}$  & $\num{0.49397}$ & $\num{5.5723}$\\
  & MLEM & $\num{0.71033}$ & $\num{0.0625}$  & $\num{0.11765}$ & $\num{3.1285}$\\
\hline
\end{tabular}
}
\label{tab:results}
\end{table}


\begin{figure}
\centering 
\begin{subfigure}{0.16\textwidth}
    \includegraphics[width=\textwidth]{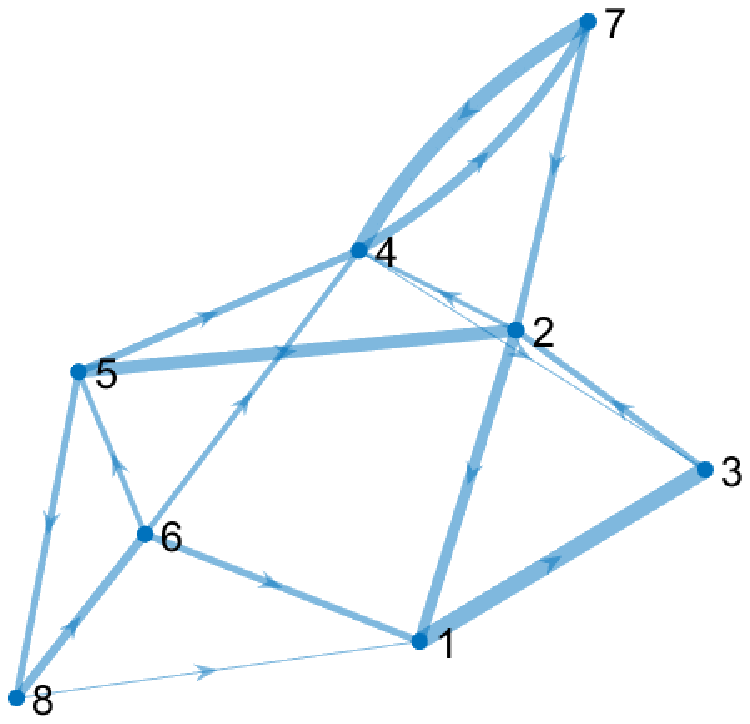}
\end{subfigure}
\hfill
\begin{subfigure}{0.15\textwidth}
    \includegraphics[width=\textwidth]{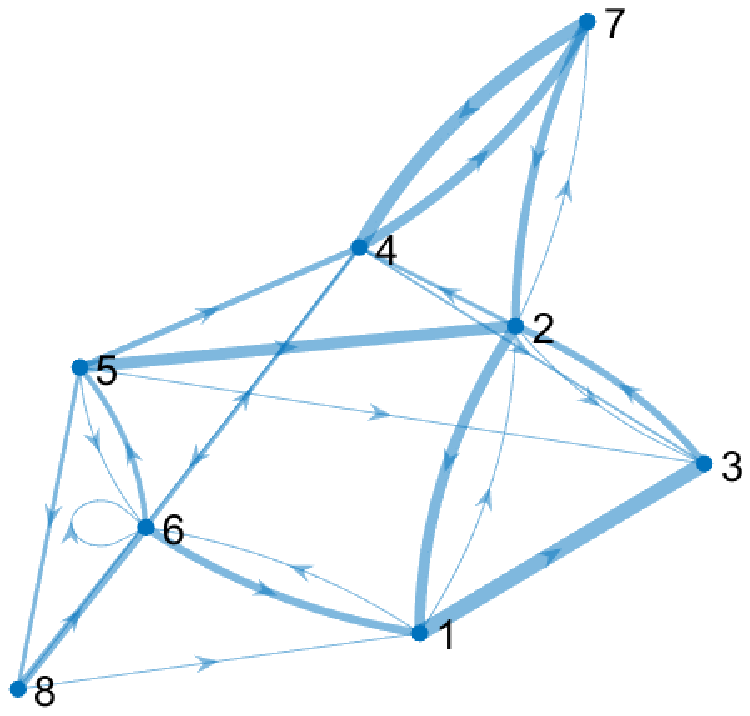}
\end{subfigure}
\hfill
\begin{subfigure}{0.16\textwidth}
    \includegraphics[width=\textwidth]{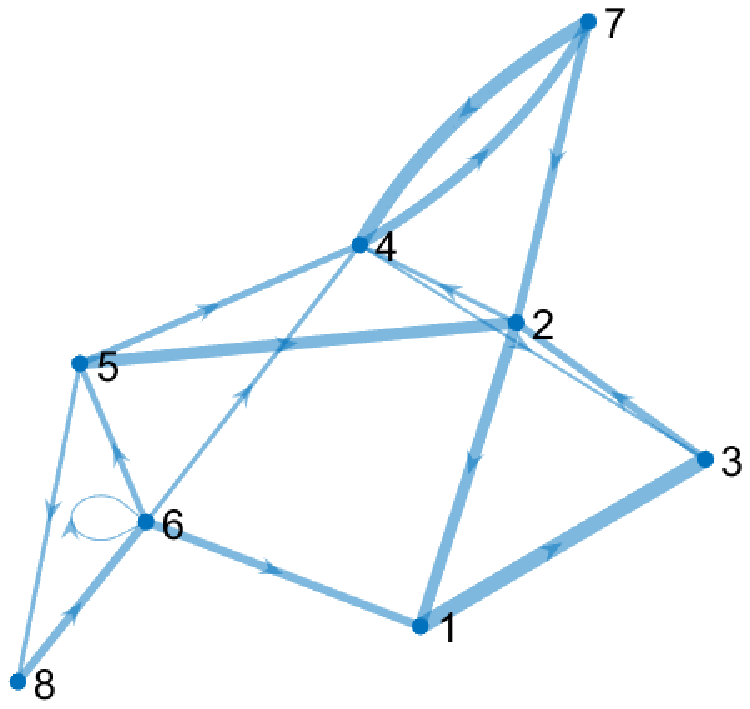}
\end{subfigure}
\begin{subfigure}{0.16\textwidth}
    \includegraphics[width=\textwidth]{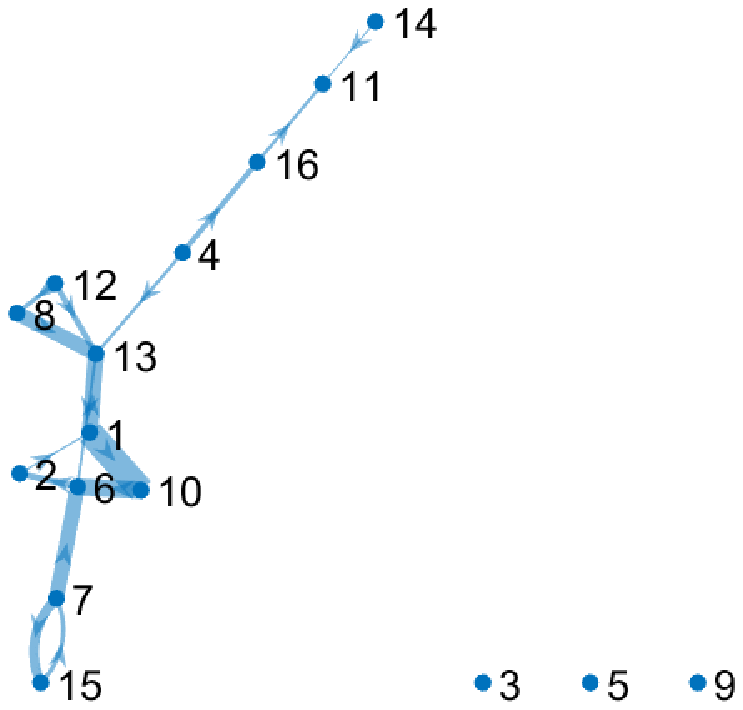}
\end{subfigure}
\hfill
\begin{subfigure}{0.15\textwidth}
    \includegraphics[width=\textwidth]{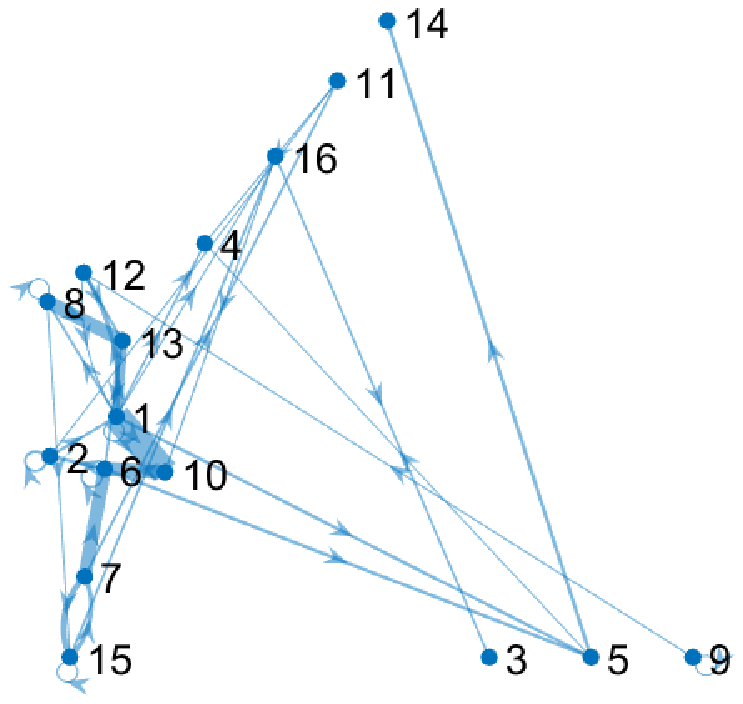}
\end{subfigure}
\hfill
\begin{subfigure}{0.16\textwidth}
    \includegraphics[width=\textwidth]{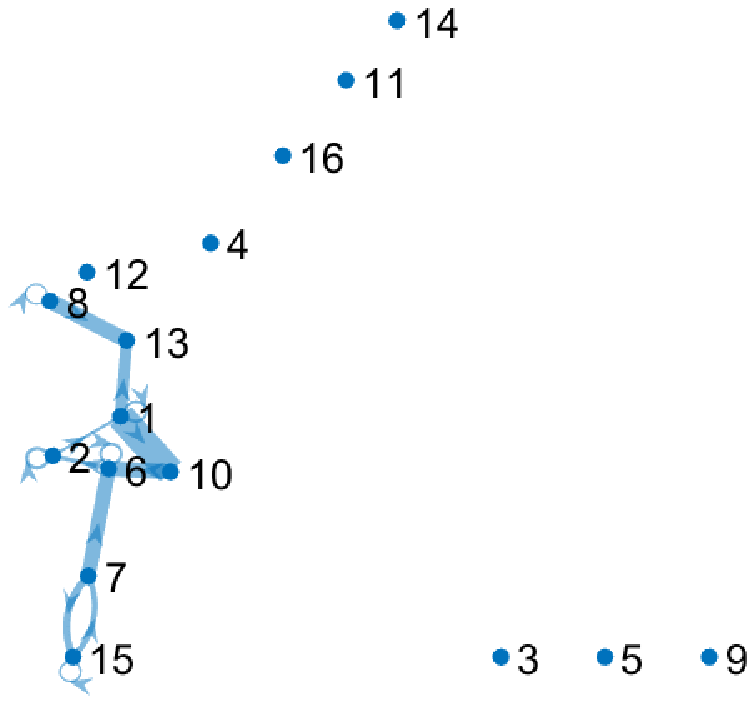}
\end{subfigure}        
\caption{True graph (left), GraphEM (middle), and GraphIT (right) estimates for $(N_x,S) = (8,16)$ (top) and $(N_x,S) = (16,16)$ (bottom).}
\vspace*{-0.2cm}
\label{fig:datasetC}
\end{figure}


\section{CONCLUSION}
\label{sec:conclusion}
In this paper, we have proposed GraphIT, an MM algorithm for the estimation of a sparse transition matrix describing hidden state interactions in an LG-SSM. A family of non-convex regularization term is considered to enforce the sparsity of the sought matrix, here interpreted as related to the adjacency matrix of a directed graph. The novel method exploits majorization properties inherited from both the EM framework and the iterated reweighted $\ell_1$ methodology. The resulting convex upper bounds can be efficiently minimized through a proximal splitting solver. Through numerical results on controlled scenarios, we illustrate the great ability of the method to properly retrieve sparse transition matrices.


\bibliographystyle{IEEEbib_short}
\scriptsize
\bibliography{references}
\end{document}